\documentclass[a4paper]{article}

\usepackage{fancyhdr}
\input{epsf}

\def\ct{\cite}
\def\be{\begin{equation}}
\def\ee{\end{equation}}
\def\bi{\bibitem}

\begin{document}
\date{July 29, 2001}
\title{Reduced total energy requirements for a modified Alcubierre warp drive spacetime}

\author{F. Loup\\{\em Royal Sun Alliance}\\ Avenida Duque De Avila 141 2 Andar 1050 Lisbon Portugal\footnote{This address is given for mailing purposes only, although I hold a professional position at the above establishment my work was carried out independently and does not reflect RSA.; email: loupwarp@yahoo.com}\and
D. Waite\\{\em Modern Relativity}\footnote{homepage:
http://www.modernrelativity.com; email: FineS137@aol.com}\and E.
Halerewicz, Jr.\\{\em Lincoln Land Community College}\\ 5250
Shepherd Road, Springfield, IL 62794 USA\footnote{Student of
above institution only; email: ehj@warpnet.net}}
\maketitle

\begin{abstract}
It can be shown that negative energy requirements within the Alcubierre spacetime can be greatly reduced when one introduces a lapse function into the Einstein tensor.  Thereby reducing the negative energy requirements of the warp drive spacetime arbitrarily as a function of $A(ct,\rho)$.  With this function new quantum inequality restrictions are investigated in a general form.  Finally a pseudo method for controlling a warp bubble at a velocity greater than that of light is presented.
\end{abstract}

\section{Introduction}
In recent years the possibility of interstellar travel within a human lifetime has become a hurtle for theoretical physics to over come.  The discussion involves the use of radically transforming the geometry of spacetime to act as a global means of providing apparent Faster Than Light (FTL) travel.  In 1994 Miguel Alcubierre then of Wales University introduced an arbitrary spacetime function which provided an apparent means of FTL travel within the frame work of General Relativity (GR) \ct{Alc}.  The drawbacks of this revolutionary new form of propulsion results from the production of causally disconnected spacetime regions and major violations of  the standard energy conditions within GR.  Further investigations into the warp drive spacetime have suggested that any manipulation of the spacetime coordinates in regards to FTL travel requires negative energy densities \ct{Olu}.   Alcubierre's generic model coupled with causally disconnected spacetime regions and large negative energy densities make it hard to build a justifiable model of a working warp drive spacetime.  However recently it has been shown that quantum inequalities can allow for the existence of negative energy densities \ct{For}, and it has also been found that classical scalar fields can generate large fluxes of negative energy in comparison to Quantum Inequality (QI) restrictions \ct{F-R}.  Coupled with recent astrophysical and cosmological observations in support of negative energy in the form of Quintessence \ct{Per} we feel that there is still new life to be brought into the warp drive spacetime.  The once distant dream of reaching the stars may become a reality with the aid of the warp drive spacetime using latin we can express our motto as Ex Somnium Ad Astra (ESAA) which translates; from a dream to the stars\footnote{The authors would like to note that the ESAA motto was created by Simon Jenks who helped to formalize the early discussions regarding this work.}.  The following work was made possible as a collaboration of the ESAA motto in order to produce a physically justifiable model of the warp drive spacetime\footnote{The collaboration of this work was made possible through the online discussions forum; {\em Alcubierre Warp Drive} currently available at the following URL:\\ http://clubs.yahoo.com/clubs/alcubierrewarpdrive}.  The inspiration behind this work was to choose a somewhat less arbitrary means of defining the \textit{Alcubierre Warp Drive} spacetime.  Specifically we have set out the task of redefining the functions of the warp drive spacetime in order to reduce the overall energy requirements as seen in respect to the Einstein tensor.

\subsection{The Alcubierre spacetime}
The Story of the Warp Drive begins with the Alcubierre paper \ct{Alc},
where we have the following metric:
\be
ds^2=dt^2-(dx-v_s f(r_s)dt)^2-dy^2-dz^2. \label{a} \ee We have
also modified the metric signature in (\ref{a}) to correspond to
$(1,-1,-1,-1)$, for reasons which will become clear in later
sections.  The Alcubierre spacetime is further defined with the
following functions
\be
v_s(t)={dx_s(t)\over dt}
\ee
\be
r_s(t)=[(x-x_s(t))^2+y^2+z^2]^{1/2}. \ee This spacetime model is
often referred to a `top hat' model, by means of a bump parameter
$\sigma$, which is seen through
\be
f(r_s)={\tanh(\sigma(r_s +R))-\tanh(\sigma(r_s -R))\over
2\tanh(\sigma R)}\label{migfunc} \ee This equation then allows for a ``warp bubble" to develop in which a {\it star ship} may ride.  The generalities of the Alcubierre top hat function can be described by the given volume expansion
\be
\theta=v_s {x_s \over r_s}{df\over dr_s}\label{trigtoph}. \ee

This is where we receive our first scientific definition of a `warp drive.' \textit{The volume elements expand behind the spaceship, while contracting in front of it}.  The draw back of this work occurs with equation 19, where the energy density $T^{ab}n_a n_b$ violates the weak, strong, and dominate energy conditions as seen by an Eulerian observer \ct{Alc}:
\be
\rho=T^{ab}n_a n_b=a^2\;T^{00}={c^2\over 8\pi G}\;G^{00}=-{c^2\over 8\pi G}{v^{2}_{s}\rho ^2 \over 4r_s^{2}}\left({df\over dr_s }\right)^2.
\ee

\section{The ESAA spacetime}
One can choose to describe the function of the warp drive in relationship to the function $g(\rho)$, where $\rho=r_s$ and $g(\rho)=1-f(\rho)$.  We approach this aspect in a manner not to dissimilar to the one chosen by Hiscock \ct{His} where we impose a pseudo metric transformation through the following functions
\begin{eqnarray}
g_{00}={A(\rho)^2-[v_s(r)\;g(\rho)]^2}\label{wfun1}\\
g_{01}=g_{10}=-v_s\;g(\rho)\\
g_{11}=g_{22}=g_{33}=-1 \label{wfun3}
\end{eqnarray}
In order to reduce the energy requirements arbitrarily we consider
the following lapse function $A(ct,\rho)$.  Where have set $A=1$
both at the location of the ship, and far from it, but allow it to
become large in the warped region following the lapse function
$A(ct,\rho)$.  Such that the above functions
(\ref{wfun1}-\ref{wfun3}) have the following metric signature in
cylindrical coordinates:
\be
ds^2=[(A)-{v_s(r)\;g(\rho)}]^2+2v_s g(\rho)dctdz'-dz'^2-dr^2-r^2d\phi^2.\label{waite1}
\ee We further define (\ref{waite1}) with the following coordinate transformation:
\be
z'=z-\int^{ct}v_s dct\label{prime} \ee such that the ship velocity
has the following definition $v_s=1=c$ thereby arriving at
\begin{eqnarray}
ds^2=&(A(ct,\rho)^2-v_s(ct)^2g(\rho)^2)dct^2-2v_s(ct)g(\rho)\cos(\theta)dctdr\nonumber\\+&2v_s(ct)g(\rho)\sin(\theta)\rho\;dctd\theta-dr^2-\rho^2d\theta^2-\rho^2\sin^2(\theta)d\phi^2.
\label{waite2}
\end{eqnarray}
Thus the exact the solution for the Einstein tensor $G^{00}$ for ({\ref{waite2}) is given by
\begin{eqnarray}
G^{ct\;ct}=&-{1\over 4}v_s(ct)^2\left({\partial\over\partial\rho}\;g(\rho)\right)\sin^2(\theta)/\bigl(A(ct,\rho)^2-v_s(ct)^2 g(\rho)^2\nonumber\\+&v_s(ct)^2g(\rho)^2\cos^2(\theta)+v_s(ct)^2g(\rho)^2\sin^2(\theta)\bigr)^2
\end{eqnarray}
where we have imposed the trigonometric identity $sin^2(\theta)+cos^2(\theta)=1$.

We now just briefly discuss how the dramatic energy reductions can occur within the warp drive spacetime.  First beginning with the Alcubierre metric in the ship frame coordinates which is stated so that the function of interest is $g(\rho)$ instead of $f(\rho)$.  Alcubierre originally introduced a gravitational time dilation term that doesn't limit ship speed in the $g_{00}$ term of his metric that has been overlooked thus far.  Instead of inserting this term starting with the remote frame metric as Alcubierre did, we have inserted it into the ship frame metric. With these coordinates an exact calculation of the entire stress energy tensor can be made fairly easy. The $T^{00}$ term is the ship frame energy density term debated to be to large. Its solution with the lapse function $A(ct,\rho)$ inserted is
\be
T^{00}=-{v_s\over 4}{c^4\over 8\pi G}\left({dg\over
d\rho}\right)^2{\{\sin(\theta)\}^2\over
\{(A(ct,\rho)\}^4}\label{waitetensor} \ee
see section \ref{TEC} to see how this equation was derived.  Letting $A$ become large arbitrarily reduces this negative ship frame energy density requirement.  Due to the non vanishing presence of the other stress-energy tensor terms there will still be negative energy in other frames, and thus a Weak Energy Condition (WEC) violation, but we really don't know that the ``quantum implied" weak energy condition is a frame covariant principle even though it has been expressed in a frame covariant manner. [$T_{ab}U^aU^b\geq 0$ for all time like $U^a$.] Since the ship forms the warp, satisfying that there is very little ship frame negative energy may be enough.

\section{Quantum Inequalities}\label{QI}
Pfenning applied \ct{Pfe} a quantum inequality for a free, massless scalar field to the Alcubierre warp even though the Alcubierre spacetime is not the result of a free, massless scalar field. Because of this, the results for the restriction on the warp shell's thickness are unreliable. It is this quantum inequalities restriction on the warp shell's thickness that causes the negative energy's magnitude to be so great. Therefore this result is also unreliable. Pfenning also did not include the possibility of an $A(ct,\rho)$ term other than $A=1$. Even though the quantum inequality may not be reliably applicable to the warp drive spacetime we can go back and redo the calculation with a variable $A(ct,\rho)$ included. The quantum inequality is
 \be
{\tau_0\over\pi}\int_{-\infty}^{+\infty}{\left<T^{\mu\nu}U_\mu U_\nu
\right>\over\tau^2+\tau_0^2}d\tau\geq{h\over c}{3\over 32\pi^2\tau_0^4}\label{qeq1}
\ee
In order to apply it we must find $T^{\mu\nu}U^\mu U^\nu$ and $\tau_0$ for a Eulerian observer. A Eulerian observer is an observer who starts out just inside the warp shell at its equator with zero initial velocity with a \textit{small sampling time} according to the ship frame who ``samples" the time it takes $\tau_0$ for the negative energy region to pass him.

From the above interval we have eq. (\ref{waite1})
\be
c^2=c^2(A^2-v_S^2g^2)(dt/d\tau)^2-2v_s gc(dt/d\tau)U-U^2
\ee
thus from the definition of the lapse function we have
\be
{dt\over d\tau}=A^{-1}. \ee Combining these (with a minimum
sampling time) results in
\be
0={1\over 2}U^2+v_s gcA^{-1}U+ {1\over 2}v_s^2g^2A^{-2}
\ee
\be
U = -v_s gcA^{-1}
\ee
\be
[U^\mu]=\left[{\begin{array}{*{20}c}
{cA^{ - 1}}\\
{-v_s gcA^{-1}}\\
0\\
0\\
\end{array}}\right]
\ee
\be
[U_\mu]=[\begin{array}{*{20}c}
   {cA} & 0 & 0 & 0  \\
\end{array}].
\ee
By inserting $T^{\mu\nu}U_\mu U_\nu=A^2T^{00}c^2$ into the inequality (\ref{qeq1}) we have
\be
{\tau_0\over\pi}\int_{-\infty}^{+\infty}{A^2T^{00}\over\tau^2+\tau_0^2}d\tau\geq{h\over c^3}{3\over 32\pi^2\tau_0^4}
\ee
and
\be
\tau_0\int_{-\infty}^{+\infty}{\beta^2\over A^2\rho^2}\left({dg\over d\rho}\right) ^2{d\tau\over\tau^2+\tau_0^2}\leq{Gh\over c^7}{3\over r^2\tau_0^4}
\ee
where we can make the replacement
\be
\rho^2=r^2+z^2
\ee
with $r=const$.

Pfenning then makes an approximation of the geodesic motion of the Eulerian observer. Equation eq. 5.11 (note further references to Pfenning in this section refer to \ct{Pfe}).  Since we are using the ship frame with $A(ct,\rho)$ inserted, we must include its effect here as well
\be
z\approx -vg(r)A^{-1}(r)\tau
\ee
\be
\rho^2=r^2+v^2g^2A^{-2}\tau^2
\ee
such that
\be
\tau_0\int_{-\infty}^{+\infty}{v^2\over A^2r^2+v^2g^2\tau^2}\left({dg\over d\rho}\right)^2{d\tau\over\tau^2+\tau_0^2}\leq{G\hbar\over c^5}{3\over r^2\tau_0^4}.
\ee
We make a definition corresponding to Pfenning's equation 5.15
\be
\psi={r\over [vg(r)]}
\ee

And according to Pefenning 5.16 the integral should be approximately
\be
{\pi\over A\psi(\tau_0+A\psi)}\leq{G\hbar\over c^5}{3\Delta^2\over v^2\psi^2\tau_0^4}
\ee
So the inequality becomes
\be
{\pi\over 3}\leq{G\hbar\over c^5}{\Delta^2[A(r)]^2\over
v^2\tau_0^4}\left({v\tau_0\over rA(r)}g(r)+1\right). \ee This
result is the same as Pfenning eq 5.17 with the exception that
$A(r)$ is not necessarily 1. Pfenning then asserts that this
result must hold for sample times that are small compared to the
square root inverse of the largest magnitude of the Riemann tensor
components as calculated for the local observers frame.

If for simplicity A is kept approximately constant through the negative energy region, then the largest Riemann tensor component for this spacetime is
\be
\left|R^\rho_{00\rho}\right|\approx \beta^2(dg/d\rho)^2 ~ (v/\Delta)^2
\ee
So in this case, the sampling time must be restricted to
\be
\tau_0=\alpha\Delta/v
\ee
where $0<\alpha <1$.

Inserting this into the inequality and looking at the case of large A approximately constant $A_0$ through the negative energy region leads to
\be
\Delta\leq (3/\pi)^{1/2}(Gh/2\pi c^3)^{1/2}(v/c)A_0/\alpha^2
\ee
Choosing $\alpha=0.10$, this approximates to
\be
\Delta\leq 10^2(v/c)l_{Pl}A_0
\ee
Now we see that letting A become arbitrarily large also arbitrarily thickens the minimum warp shell thickness. Therefore the -0.068 solar mass calculation \ct{Pfe} had a reachable shell thickness. All that remains then is to divide the Pfenning -0.068 solar mass result by an $A_0^4$ which would allow the thickness chosen, and which will lower the energy magnitude even farther by several orders of magnitude.

\section{Total Energy Calculations}\label{TEC}
The energy requirements for the warp drive spacetime can be calculated from
\be
E=\int T^{00}dV
\ee
For $T^{00}$ we have
\be
T^{00}=-{v_s c^4\over 32\pi G}\left({dg\over d\rho}\right)^2{r^2\over r^2+z^2}{1\over A^4}
\ee
Where we assume $A=const$ to be large, one can also see that this is identical to eq. (\ref{waitetensor}).  We now choose
\[
dV=\rho^2\sin(\theta)d\rho d\theta d\phi
\]
so that we can write
\be
E=-v_s^2{c^4\over 12G}\int_0^\infty\rho^2\left({dg\over d\rho}\right)^2\left(\frac{1}{A^4}\right)d\rho
\ee
which reduces to
\be
E_{ESAA}=-\int_0^\infty{v_s\over 12}{c^4\over G}\left({dg\over
d\rho}\right)^2\left({1\over A^4}\right)\rho^2d\rho\label{tec1}. \ee Thus from section \ref{QI} we can show similarly that the energy can be given from:
\be
E=-\int_0^\infty{v_s^2c^4\over
12G}\left({1\over\Delta}\right)^2\left({1\over
A^4}\right)\rho^2d\rho.
\ee

\section{Post relativistic warp drives?}\label{hyper}
One of the requirements for a warp drive spacetime which has a
velocity greater than that of light is the existence of negative
energy \ct{Olu}.  However this negative energy, exotic matter,
quintessence, or however you chose to define it can be made to
pseudoly appear from negative brane tension \ct{brane}.  More
specifically for our case, an Anti de Sitter (AdS) spacetime would
b e a logical choice for such an exploration.  The reason for this
choice is the fact that $\Lambda$ acts as a scalar field to
exploit negative energy densities \ct{F-R}. In reference to
\ct{glv} we can write an arbitrary metric for a single ``warped"
extra dimension that has 3-D rotational invariance in AdS space by
\be
ds^2=-a^2(r,t)dt^2+b^2(r,t)d\vec x^2+c^2(r,t)dr^2 \label{aAdS} \ee
This equation at a first approximation is identical to eq.
(\ref{waite1}).  Now considering eq. (\ref{waite2}) in reference
to eq. (\ref{aAdS}) a five-dimensional AdS warp drive spacetime
can be given by
\begin{eqnarray}
d\hat s^2=&-\Lambda_{b\kappa}g(\hat\rho)^2dt^2+A(c\hat t,\hat\rho)d \Sigma_\kappa^2+(A(c\hat t,\hat\rho)-v_s(c\hat t,\hat\rho)^2g(\hat\rho)^2)\nonumber\\&dc\hat t^2+d\hat\rho^2+\hat\rho^2d\theta^2+\hat\rho\sin^2\theta d\phi^2
\label{5dwarp}
\end{eqnarray}
where
\be
d\Sigma^2_\kappa={d\sigma^2\over (1-\kappa
L^{-2}\sigma^2)}+\sigma^2 d\Omega^2_2 \ee  with $\kappa$ being the
curvature scale and L being the length scale of the bulk
$\Lambda_{b\kappa}$.  Where $ct,\rho=const$ and $d\Sigma$ is a
unit metric for a 3 brane.  From this the Alcubierre warp drive
function $f(\rho)$ expands and contracts within the (3+1) slice of the
AdS spacetime via a scale factor dependent on time $(c\hat t,0)$.
Since the five-dimensional AdS spacetime contains slices of the
four-dimensional space the time coordinates rescale differently at different points in the extra dimension of the de Sitter space.  Thus
causing the $(ct,\rho)$ terms to scale differently in the 5-D
space than it does in the 4-D space, therefore a gravitational
wave in 4-D space can appear to exceed the light because the bulk
spacetime acquires different velocities for $c=1$ when moving
through the `extra' dimension.  This spacetime would also act to
suppress the ``need one to make one" paradox of \ct{Cou}.  When
the 5-D space is static and there is no gauge field one can have a
more general solution
\be
ds^2=-A(ct,\rho)dt^2+L^{-2}\rho^2d\Sigma_\kappa^2+(A(ct,\rho)-v_s(ct,\rho)^2g(\rho)^2)d\rho^2\label{adsbro}
\ee
with
\be
L^{-2}=-{1\over 6}\kappa_5^2\Lambda_{bk} \ee which would
essentially describe a warp drive in which $v_0<c$.  What is
interesting to note about eq. (\ref{5dwarp}) is that it can make
the 4-D warp drive function $g(\rho)$, appear to be generated from
a cosmologic term of order $\Lambda g(\rho)$ in AdS space. Meaning
that matter within the warp bubble wouldn't need to become
tachyonic to solve the horizon [control] problem of the Alcubierre
Warp Drive as $c\leq v_0$ (i.e. the scalar field acts to create a
hyper warp drive without large negative energy requirements), it
is however noted that inducing the fifth dimension is purely a
mathematical trick which may have problems of its own.  In the
tradition of Alcubierre the warp drive of eq. (\ref{5dwarp}) just
begs  to be named after its familiar counterpart, the
``hyperdrive" of science fiction.  We also note that a similar but
more conventional approach was taken by Gonz\'alez-Di\'az
\ct{Gon}, in which a Alcubierre-Hiscock spacetime is embedded in a
three-dimensional Misner space, which fits into the frame work of
classical GR as opposed to this discussion.

\subsection{a Broeck extended AdS spacetime?}\label{adsbs}
Equation (\ref{5dwarp}), shouldn't look to surprising at first
hand, let us look at the Broeck spacetime \ct{Bro}:
\be
ds^2=-dt^2+B^2(\rho)[dx-v_s(t)f(\rho)dt)^2+dy^2+dz^2] \ee We see
that $L^{-2}\rho^2\equiv B^2(\rho)$, thus (\ref{adsbro}) can be
interpreted as an AdS Broeck spacetime, this is because $B^2(\rho)$
would naturally arise from primed coordinate transformations
\ct{Eve}, eq.(\ref{prime}) is a generic example of this.  Thus
when comparing equation 5 of \ct{Bro} to (\ref{adsbro}), something
quite odd occurs to the lapse function $A(ct,\rho)$
when $v_0>c$ effectively increasing the apparent amount of negative energy
within the local brane.  Moreover a similar scenario for
(\ref{adsbro}) implies a coefficient of order
$A^2\cdot\Lambda_{bk}g(\hat\rho)^2$. The resulting total energy of
the bulk spacetime in comparison to (\ref{tec1}) when $v_0\geq c$
is thus
\be
E_{bk}=\int_0^\infty {v_sc^4\over 12 G}{1\over A(c\hat t,\rho)^2}\;\rho^2 d\rho^3\label{adstec}
\ee
Notice that the bulk comsomologic term $\Lambda_{bk}$ acts to flip the
negative energy our local brane.  This can be expected because a
brane with negative tension can contribute and act positively
within a higher dimensional brane \ct{brane}, and therefore appear
to violate the NEC (Null Energy Condition).  This also suggest
that $\rho\equiv k_{\mu\nu}=K_{\mu\nu}^+-K_{\mu\nu}^-$, i.e. the center of the Broeck bubble contains a gravitational kink.

We now want to discuss a specific boundary condition to illustrate
the properties of a \textit{hyperdrive}.  To begin this discussion the lapse function will be given as
\be
A(ct,\rho)=\alpha_{PWL}=f_{PWL} \ee thus to arrive at specific
boundary condition we replace (\ref{migfunc}) with a function that
resembles eq. 4 of \ct{For} therefore we have the following
piecewise function:
\be
f_{PWL}(\rho)=\left\{ \begin{array}{*{20}c} {1} &
{\rho<R-{\Delta\over 2}} \\ {(1+[(\rho -R-{\Delta\over 2})(\rho
-R+{\Delta\over 2})]/D)^N} & {R-{\Delta\over
2}<\rho<R+{\Delta\over 2}} \\ {1} & {\rho>R+{\Delta\over 2}} \\
\end{array} \right.\label{pbound}
\ee It is also noted that the $R\pm(\Delta /2)$ term could be a
boundary equal to or greater than one corresponding to some value
$\alpha \geq 1$ in standard units) and that this value was chosen
arbitrarily, however for this instance we have defined it in
manner which coincides with Pfenning's analysis throughout this
work.  $N$ is the warp parameter which describes the magnitude of
the warp factor:
\be
N(\rho)_{warp}=B(\rho)=1+\alpha
\ee
this definition describes how the Alcubierre spacetime differs from our own\footnote{Where $B(\rho)$ is derived from eq. 5 of \ct{Bro} and the warp parameter $N$ we can also define alpha as $\alpha=N-1$.\label{alphN}}.

\begin{figure}[ht]
\centerline{\epsfbox{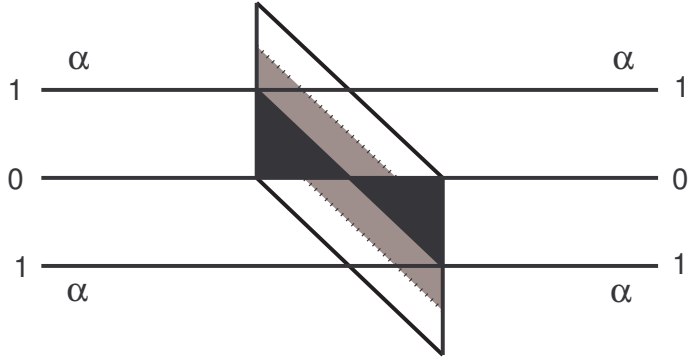}}\caption{The ESAA spacetime
represented one dimensionally in Cartesian coordinates, with
$\alpha_{PWL}=1$ this spacetime appears hyper-dimensional in
comparison to the \textit{Alcubierre spacetime} $f(\rho)=0$.  This
figure represents the varying warp parameter $N$ the minimum value
$N=B(\rho)=1$ gives the Alcubierre top hat $f(\rho)=1$ (black).  The
median (gray) and maximum (white) represents an energy reduction
due to a decreased volume as a factor of the warped region.  For
an arbitrary boundary condition $\alpha_{PWL}=(1+[(\rho
-R-{\Delta\over 2})(\rho -R+{\Delta\over 2})]/D)^N$, when
$R=100\;m$ the ``maximum warp" factor is $N=9.3$, where the lapse function becomes singular.}\label{fig1}
\end{figure}

With the boundary condition (\ref{pbound}) the energy requirements
for this warp drive spacetime can be given from:
\begin{eqnarray}
T^{\alpha\beta}n_\alpha n_\beta=&\alpha_{PWL}^2\;T^{00}\nonumber\\=&-(R+{\Delta\over 2})^{2N}\;G^{00}\nonumber\\=&
-\left[(R+{\Delta\over 2})^{2N}\right]\left[{v_s^2/4}\right][sin^2\theta]\left[{dg(\rho)/ d\rho}\right]^2
\end{eqnarray}

With $N=1$ eq. (\ref{adstec}) becomes
\be
E_{bk}={v_s c^4\over 144 G}{1\over(\hat R+\frac{1}{2}\hat\Delta)^{2N}}\int_0^\infty\rho^4 d\rho\label{finbk}
\ee
or
\be
{E_{bk}\over c^2}\approx 1.7\times 10^{38}g=1.2\times 10^5\cdot
M_{Sun}. \ee

The ESAA total energy $E_{ESAA}$ is given with $N=1$ since $v_0=c$, this depicts an \textit{Alcubierre like} spacetime from footnote \ref{alphN} we see that
\[
N=\int_1^\infty \mathop{\lim }\limits_{\alpha  \to 0} \alpha +1=0+1=1.
\]
However, we now have $v_0>c$, thus $\alpha\geq 1$ in standard units, by comparing eq. (\ref{tec1}) to eq. (\ref{finbk}) we see that $\alpha=1$, therefore meaning the hyperdrive is defined through the following parameter
\[
N=\int_1^\infty \alpha+1=1+1=2.
\]
Thus when applying an arbitrary boundary condition
$\alpha_{PWL}=(b)^N$ for the warped region, we see that $N$, would
reduce the energy requirements for an arbitrary total energy
calculation such as (\ref{tec1}).  Thus it is seen that under these
conditions the velocity of the warp bubble is to be implicitly
interpreted as $v_0=10^2c$.

In other words using a Pfenning inspired boundary condition it
takes just under one earth mass to achieve a velocity 100 times
that of light.  From the ship's frame the apparent velocity is
much less from \ct{Eve}:
\be
\frac{dx'}{dt'}={dx/dt\beta\over 1-\beta dx/dt} \ee we see that a fifteen day trip to alpha centauri, would give a local apparent velocity of
$v=0.01002c$, thus compounding the aberration effects of
\ct{null}.

\subsection{metric patching}
From this it can be seen why the warp drive violates the standard
energy conditions in GR, assuming a negative brane tension one
has \ct{brane}:
\be
T^{\mu\nu}k_\mu k_\nu=-\Lambda_D g_{induced}^{\mu\nu}k_\mu k_\nu\delta^{n-p}(n^a)=\Lambda_D\left[\sum\limits_{a=1}^{n-p}{(n_a^\mu k_\mu)^2}\right]\delta^{n-p}(n^a)\label{bcor}
\ee
from this the NEC becomes $T^{\mu\nu}k_{\mu}k_{\nu}<0$.  When a killing vector $K^\mu$ is inserted in (\ref{bcor}), for the right hand side we have
\be
T^{\mu\nu}k_\mu k_\nu K^\mu=\alpha
T^{\mu\nu}=\Lambda_D\left[\sum\limits_{a=1}^{n-p}{(n_a^\mu
)^2}\right]\delta^{n-p}(n^a) \ee which is exactly what we would
expect from (\ref{5dwarp}), with the presence of a gauge field.
In terms of the AdS geometry this is caused by the scale factor
$(c\hat t,\rho)$, such that the time lapse function is given
by the relation $A^2\Lambda_{bk}=A(c\hat t,\rho)^2$, i.e.
converting the brane negative energy into positive energy in the
bulk frame, as suggested above.

The total energy in the bulk is positive with
$r_s\in[\infty,a(t)]$ thus with metric patching the energy in our
three brane can be calculated from:
\be
G^{\mu\nu}=\eta^{\mu\nu}+\hbar^{\mu\nu}+K_-^{\mu\nu}-K_+^{\mu\nu}
\ee Throughout our analysis of the \textit{hyperdrive} we have
considered a metric patching of (3+1) within an embedding in
$E^4$, while in previous works the warp drive has been considered
in the form (2+1) embedded in $E^3$ \ct{Gon}.  A similar
construction has also been considered for transversible wormholes
\ct{brane} in reference to string theory, which conforms to an
(n+1) bulk and an ([n-1]+1) brane, again as suggest by Everett the
warp drive appears as a special case wormhole (with the exception
that this wormhole appears to be a shortcut through
\textit{hyperspace}).

\section{Summary}
The goal we set out was to reduce the energy requirements of the
warp drive spacetime to more physically realistic amounts.  Our
analysis has shown that simply considering a warp drive spacetime
with an arbitrary lapse function can dramatically lower the energy
requirements within the Einstein tensor.  We have shown that
Quantum Inequalities are a poor choice for `curved spacetimes,'
but even so minor modifications of QI's can also reduce their
allowable energy restrictions.  Finally we have derived a
mathematical trick which may allow the warp drive spacetime to
survive within the superluminal range.  In closing it can be shown
that the warp drive spacetime can not be ruled out because of
`unphysical energy conditions' alone, by simply choosing an
arbitrary function $A(ct,\rho)$ the energy reductions become
rather dramatic.  It is however noted that such an arbitrary
function is merely a mathematical devise, within a
four-dimensional spacetime there is no mechanism to set $N>1$ as
far as we know.  However, it is somewhat easy to consider $N=2$
for a possible ``hyperdrive" spacetime, with our discussion this
would correspond to a  five-dimensional AdS spacetime (in which a
``warp bubble" could be interpreted to move with a velocity that
is 100 times that of light), and without experimental verification
of the fifth coordinate this may also be considered a clever trick
of its own.  It is therefore noted that the warp drive spacetime
appears to have several limitations within semi-classical GR, e.g.
the violation of the Null Energy Condition (NEC), therefore it is
likely that a more plausible warp drive may reside within post
relativistic corrections of GR, indeed \textit{Ex Somnium Ad
Astra}.

\section*{Acknowledgement}
The authors would like to express their sincere gratitude towards
Miguel Alcubierre for his helpful discussions and appreciated
comments throughout the developments of this work.

\appendix

\section{Selected exact contrivariant solutions of the Einstein tensor}
The exact solutions to the ESAA metric (\ref{waite2}) were
produced in Maple VI with the aid of GRTensorII and are as follows
(and for that reason $v_s$ is represented by $\beta$ in the
following calculations):

\begin{eqnarray}
G^{ct\;\rho}=&{1\over 4}\Bigl(\bigl(4A(ct,\rho)^2+4\beta(ct)2g(\rho)^2\cos^2(\theta)-4\beta(ct)^2g(\rho)^2\nonumber\\+&\beta(ct)^2g(\rho)\sin^2(\theta)\bigr)\left({\partial\over\partial\rho}g(\rho)\right)\rho\nonumber\\+&4\beta(ct)^2g(\rho)\sin^2(\theta)\left({\partial\over\partial\rho}g(\rho)\right)\cos(\theta)\beta(ct)\Bigr)\nonumber\\/&\rho\left(A(ct,\rho)^2-\beta(ct)^2g(\rho)^2\cos^2(\theta)+\beta(ct)^2g(\rho)^2\sin^2(\theta)\right)^2
\end{eqnarray}
\begin{eqnarray}
G^{\rho\;\theta}=&{1\over 4}\Bigl(\bigl(2A(ct,\rho)^2\rho\left({\partial\over\partial ct}\beta(ct)\right)+4A(ct,\rho)^2\beta(ct)^2g(\rho)\cos(\theta)\nonumber\\-&2\beta(ct)\rho\left({\partial\over\partial ct}A(ct,\rho)\right)A(ct,\rho)+\beta(ct)^4g(\rho)^2\nonumber\\&\cos(\theta)\sin^2(\theta)\left({\partial\over\partial\rho}g(\rho)\right)\rho\nonumber\\+&4\beta(ct)^4g(\rho)^3\cos(\theta)^3-4\beta(ct)^4g(\rho)^3\nonumber\\&\cos(\theta)+4\beta(ct)^4g(\rho)^3\cos(\theta)\sin^2(\theta)\bigr)\nonumber\\&\left({\partial\over\partial\rho}g(\rho)\right)\sin(\theta)\Bigr)\bigl(\rho^2\bigl(A(ct,rho)^2-\beta(ct)^2g(\rho)^2\nonumber\\+&\beta(ct)^2g(\rho)^2\cos^2)(\theta)+\beta(ct)^2g(\rho)^2\sin^2(\theta)\bigr)^2\Bigr)
\end{eqnarray}

\section{Selected exact covariant solutions of the Einstein tensor}
Again these are the solutions to the metric (\ref{waite2}), with the $v_s$ to $\beta$ modifications:
\begin{eqnarray}
G_{\rho\rho}=&{1\over4}\bigl(8A(ct,\rho)\left({\partial\over\partial\rho}A(ct,\rho)\right)+4\beta(ct)^2g(\rho)\sin^2(\theta)\left({\partial\over\partial\rho}g(\rho)\right)^2\sin^2(\theta)\rho\nonumber\\-&8\beta(ct)^2g(\rho)\left({\partial\over\partial\rho}g(\rho)\right)\bigr)\rho\bigr(A(ct,\rho)^2-\beta(ct)^2g(\rho)^2\nonumber\\+&\beta(ct)^2g(\rho)^2\cos^2(\theta)+\beta(ct)^2g(\rho)^2\sin^2(\theta)\bigr)
\end{eqnarray}
\begin{eqnarray}
G_{ct\;ct}=&-{1\over 4}\beta(ct)^2\bigl(-4\rho A(ct,\rho)^3\left({\partial^2\over\partial\rho^2}A(ct,\rho)\right) g(\rho)^2\sin^2(\theta)\nonumber\\+&4\beta(ct)^2g(\rho)^3\sin^4(\theta)\rho\left({\partial^2\over\partial\rho^2}g(\rho)\right) A(ct,\rho)^2\nonumber\\+&4\beta(ct)^2g(\rho)^3\cos^2(\theta)\sin^2(\theta)\left({\partial^2\over\partial\rho^2}g(\rho)\right)\rho A(ct,\rho)^2\nonumber\\-&4\rho A(ct,\rho)^2\beta(ct)^2g(\rho)^3\left({\partial^2\over\partial\rho^2}g(\rho)\right)\sin^2(\theta)\nonumber\\-&4\rho\beta(ct)^2g(\rho)^4\sin^2(\theta)A(ct,\rho)\left({\partial^2\over\partial\rho^2}A(ct,\rho)\right)\cos^2(\theta)\nonumber\\+&4g(\rho)\sin^2(\theta)\rho\left({\partial^2\over\partial\rho^2}g(\rho)\right)A(ct,\rho)^4\nonumber\\+&4\beta(ct)^2g(\rho)^3\cos^2(\theta)\sin^2(\theta)\left({\partial^2\over\partial\rho^2}g(\rho)\right)\rho A(ct,\rho)^2\nonumber\\
-&4\rho A(ct,\rho)\beta(ct)^2g(\rho)^3\left({\partial\over\partial\rho} g(\rho)\right)\left({\partial\over\partial\rho}A(ct,\rho)\right)\sin^2(\theta)\nonumber\\+&3\rho\beta(ct)^4g(\rho)^4\sin^2(\theta)\left({\partial\over\partial\rho} g(\rho)\right)^2\cos^2(\theta)\nonumber\\+&4\beta(ct)^2g(\rho)^4\sin^2(\theta)A(ct,\rho)\left({\partial\over\partial\rho}A(ct,\rho)\right)\nonumber\\-&4\beta(ct)^2g(\rho)^3\sin^2(\theta)A(ct,\rho)^2\left({\partial\over\partial\rho}g(\rho)\right)\nonumber\\-&4\beta(ct)^2g(\rho)^4\sin^4(\theta)A(ct,\rho)\left({\partial\over\partial\rho}A(ct,\rho)\right)\nonumber\\-&8g(\rho)^4\cos^2(\theta)\beta(ct)^2A(ct,\rho)\left({\partial\over\partial\rho}A(ct,\rho)\right)\nonumber\\-&8g(\rho)^4\cos^4(\theta)\beta(ct)^2A(ct,\rho)\left({\partial\over\partial\rho}A(ct,\rho)\right)\nonumber\\-&12g(\rho)^4\cos^2(\theta)\beta(ct)^2A(ct,\rho)\left({\partial\over\partial\rho}A(ct,\rho)\right)\sin^2(\theta)\nonumber\\-&8g(\rho)^3\cos^2(\theta)\beta(ct)^2\left({\partial\over\partial\rho}g(\rho)\right)A(ct,\rho)^2\nonumber\\-&3\beta(ct)^2\sin^4(\theta)\left({\partial\over\partial\rho}g(\rho)\right)^2\rho A(ct, \rho)^2g(\rho)^2\nonumber\\-&3\beta(ct)^2\sin^2(\theta)\left({\partial\over\partial\rho}g(\rho)\right)^2\rho g(\rho)^2\cos^2(\theta)A(ct, \rho)^2\nonumber\\+&4\rho\beta(ct)^2g(\rho)^4\left({\partial\over\partial\rho}A(ct,\rho)\right)^2\sin^2(\theta)\nonumber\\-&3\rho\beta(ct)^4g(\rho)^4\left({\partial\over\partial\rho}g(\rho)\right)^2
\sin^2(\theta)\nonumber\\-&4\rho\beta(ct)^2g(\rho)^4\left({\partial\over\partial\rho}A(ct,\rho)\right)^2\nonumber\\+&3\rho\beta(ct)^4g(\rho)^4\sin^4(\theta)\left({\partial\over\partial\rho}g(\rho)\right)^2\nonumber\\+&6\rho A(ct,\rho)^2\beta(ct)^2\left({\partial\over\partial\rho}g(\rho)\right)^2g(\rho)^2\sin^2(\theta)\nonumber\\-&4\rho g(\rho)\sin^2A(ct,\rho)^3\left({\partial\over\partial\rho}A(ct,\rho)\right)\left({\partial\over\partial\rho}g(\rho)\right)\nonumber\\+&4\rho\beta(ct)^2g(\rho)^3\sin^2(\theta)A(ct,\rho)\left({\partial\over\partial\rho}A(ct,\rho)\right)\left({\partial\over\partial\rho}g(\rho)\right)\cos^2(\theta)\nonumber\\+&4\rho\beta(ct)^2g(\rho)^3\sin^4(\theta)A(ct,\rho)\left({\partial\over\partial\rho}A(ct,\rho)\right)\left({\partial\over\partial\rho}g(\rho)\right)\nonumber\\-&4\rho\beta(ct)^2g(\rho)^4\sin(\theta)^2\left({\partial\over\partial\rho}A(ct,\rho)\right)^2\cos^2(\theta)\nonumber\\+&8\beta(ct)^2g(\rho)^3\cos^4(\theta)\left({\partial\over\partial\rho}g(\rho)\right)A(ct,\rho)^2\nonumber\\+&12\beta(ct)^2g(\rho)^3\cos^2(\theta)\left({\partial\over\partial\rho}g(\rho)\right)A(ct,\rho)^2\sin^2(\theta)\nonumber\\+&4\beta(ct)^2g(\rho)^3\sin^4(\theta)\left({\partial\over\partial\rho}g(\rho)\right)A(ct,\rho)^2\nonumber\nonumber\\-&8g(\rho)^2\cos^2(\theta)A(ct,\rho)^3\left({\partial\over\partial\rho}A(ct,\rho)\right)\nonumber\\-&4g(\rho)^2\sin^2A(ct,\rho)^3\left({\partial\over\partial\rho}A(ct,\rho)\right)\nonumber\\+&\sin^2(\theta)\left({\partial\over\partial\rho}g(\rho)\right)^2\rho A(ct,\rho)^4\nonumber\\+&8g(\rho)\cos^2(\theta)\left({\partial\over\partial\rho}g(\rho)\right)A(ct,\rho)^4\nonumber\\+&4g(\rho)\sin^2(\theta)\left({\partial\over\partial\rho}g(\rho)\right)A(ct,\rho)^4\bigr)\nonumber\\/&\bigl(A(ct,\rho)^2-\beta(ct)^2g(\rho)^2+\beta(ct)^2g(\rho)^2\cos^2(\theta)\nonumber\\+&\beta(ct)^2g(\rho)^2\sin^2(\theta)\bigr)^2\rho
\end{eqnarray}

\bibliographystyle{plain}

\end{document}